\documentclass[conference]{IEEEtran}
\IEEEoverridecommandlockouts

\usepackage{cite}
\usepackage{amsmath,amssymb,amsfonts}
\usepackage{algorithmic}
\usepackage{graphicx}
\usepackage{subcaption}
\usepackage{textcomp}
\usepackage{xcolor}
\usepackage{booktabs, multirow, tabularx} 
\usepackage{soul}
\usepackage{url}

\def\BibTeX{{\rm B\kern-.05em{\sc i\kern-.025em b}\kern-.08em
    T\kern-.1667em\lower.7ex\hbox{E}\kern-.125emX}}
\begin{document}

\title{Exploring Smartphone-based Spectrophotometry for Nutrient Identification and Quantification\\
}


\author{\IEEEauthorblockN{Andrew Balch}
\IEEEauthorblockA{\textit{Department of Computer Science} \\
\textit{University of Virginia}\\
Charlottesville, USA \\
xxv2zh@virginia.edu}
\and
\IEEEauthorblockN{Maria A. Cardei}
\IEEEauthorblockA{\textit{Department of Computer Science} \\
\textit{University of Virginia}\\
Charlottesville, USA \\
cbr8ru@virginia.edu}
\and
\IEEEauthorblockN{Afsaneh Doryab}
\IEEEauthorblockA{\textit{Department of Computer Science} \\
\textit{University of Virginia}\\
Charlottesville, USA \\
ad4ks@virginia.edu}
}

\maketitle

\begin{abstract}
Imbalanced nutrition is a global health issue with significant downstream effects. Current methods of assessing nutrient levels face several limitations, with accessibility being a major concern. 
In this paper, we take a step towards accessibly measuring nutrient status within the body. We explore the potential of smartphone-based spectrophotometry for identifying and quantifying nutrients in a solution by building and testing two prototype devices.
We compared the prototypes and found that the limitations posed by the initial, simpler prototype were well addressed in the more portable and reliable second-generation device. With the second-generation prototype, we created and implemented a semi-automatic signal processing and analysis pipeline for analyzing absorption spectra. We thoroughly evaluated the prototypes by analyzing the effect of four different light sources and three reference spectra strategies. Results demonstrate that an LED bulb light source performed best, and all reference spectra strategies performed similarly. We then compared the second-generation prototype to a benchtop laboratory spectrophotometer to further validate the device. We applied the Beer-Lambert Law to demonstrate that our prototype is able to quantify the amount of vitamin B12 in a solution with an accuracy of up to 91.3\%. Our in-depth analyses, discussions, and results demonstrate the potential use of smartphone-based spectrophotometry as an accessible method to identify and quantify nutrients and pave the way for future developments that can apply this approach to the human body. \\

\end{abstract}

\begin{IEEEkeywords}
health sensing, mobile health, precision nutrition, micronutrients, nutrition assessment, point-of-care devices, accessibility, spectrophotometry.
\end{IEEEkeywords}



\section{Introduction}
Balanced nutrition is crucial for the growth and proper functioning of the human body. The World Health Organization (WHO) reports that individuals with proper nutrition have increased lifespans, are more likely to break cycles of poverty, and have lower risks of disease, which impacts mortality \cite{world_health_organization_nutrition_2024}. Nutrient imbalance, an excess or a deficiency of nutrients, is a global health issue with major downstream health effects. Women, children, and underserved populations, in particular, bear the greatest burden of nutrient imbalance \cite{hummell-role-2022,drake-micronutrient-2018, bruins-maximizing-2016}.  Anemia is an example of a major resulting disease; it affects 37\% of pregnant people and 40\% of children under five years of age. Moreover, nutrient deficiency is linked to a staggering 45\% of deaths in children under five years old \cite{world_health_organization_nutrition_2024}, highlighting the severity of the nutrient imbalance, which is often challenging to identify and, consequently, hard to prevent.

Methods for measuring nutrient levels present several limitations, further perpetuating the imbalance issue. A nutrient assessment often requires invasive, expensive, and in-person clinic visits, making awareness of nutritional status largely inaccessible \cite{balch-towards-2024}. Current methods are done via indirect, subjective dietary logs, in-person clinical examinations, and complex in-lab analyses on blood (e.g., liquid chromatography-coupled mass spectrometry). While valuable, these assessments are expensive, flawed, and burdensome for the patient and the clinician.



Using low-cost, convenient components, such as a smartphone, nutrient status detection can be more accessible, especially in resource-constrained environments \cite{Scherr2017}. As a step in this direction, we explore the application of smartphone-based spectrophotometry. 
Spectrophotometry is the analysis of light (primarily visible light) as it is absorbed and transmitted by a compound it passes through \cite{vo-215-2013}. The relative concentration and identity of a compound can be determined by analyzing its absorbance at specific wavelengths.

With the ultimate goal of measuring nutrient levels within the body, this work sets the stage by investigating smartphone-based spectrophotometry to quantify and identify nutrients in a solution. We utilized low-cost, DIY (do-it-yourself) materials (e.g., smartphone, 3D-printed components) to prototype a spectrophotometer for analyzing nutrient solutions. A simple, first-generation prototype demonstrated the application of smartphone-based spectrophotometry to this problem area, but the low-quality results on our test nutrients reflect its immature design. To address the limitations faced by our initial prototype, we implemented an improved second-generation prototype based on \cite{bruininks-inexpensive-2022}, modified with a cuvette holder attachment. This prototype is more portable and reliable while still being accessible and noninvasive. We formalized a signal analysis pipeline using ImageJ \cite{schneider_nih_2012} and Python to extract absorbance spectra from smartphone images over the visible range (400 to 700 nm). To comprehensively assess our setup, we compared and analyzed four distinct light sources and three reference spectra calculation methods. We evaluated our prototypes to investigate the following questions:

\begin{enumerate}
    \item[\textit{RQ1:}] Can an accessible, smartphone-based spectrophotometer be used to detect and quantify nutrients?
    \item[\textit{RQ2:}] With what light source and reference spectra calculation method does such a spectrophotometer perform best?
    \item[\textit{RQ3:}] How does the performance of our prototype compare to results from a laboratory spectrophotometer?
\end{enumerate}
The experimental results and analyses of nutrient supplement solutions provide valuable insights into how to enhance future designs to more effectively address the issue of nutrient identification and quantification. Our contributions are as follows:

\begin{enumerate}
    \item We build an accessible smartphone-based spectrophotometer with at-home materials to identify and quantify nutrients in solutions. Take-away lessons learned and limitations are discussed to inspire further device editions to eventually enable the detection of nutrient levels within the body.
    \item We thoroughly test the impact of different design parameters on performance. We compare and analyze the impact of four distinct consumer light sources and three reference spectra strategies on the design. Analyses indicate that (1) the LED (light-emitting diode) bulb light source showed the most uniform spectra, and (2) all three reference spectra computation methods performed similarly.
    \item We evaluate the prototype against a benchtop laboratory spectrophotometer (Hach DR6000). The prototype exhibited performance that approached the laboratory spectrophotometer. However, it exhibited more noise and inter-sample variance.
\end{enumerate}


\section{Background and Related Work} \label{RW}


\subsection{Spectrophotometry Overview}

\begin{figure}
    \centering
    \includegraphics[width=0.8\linewidth]{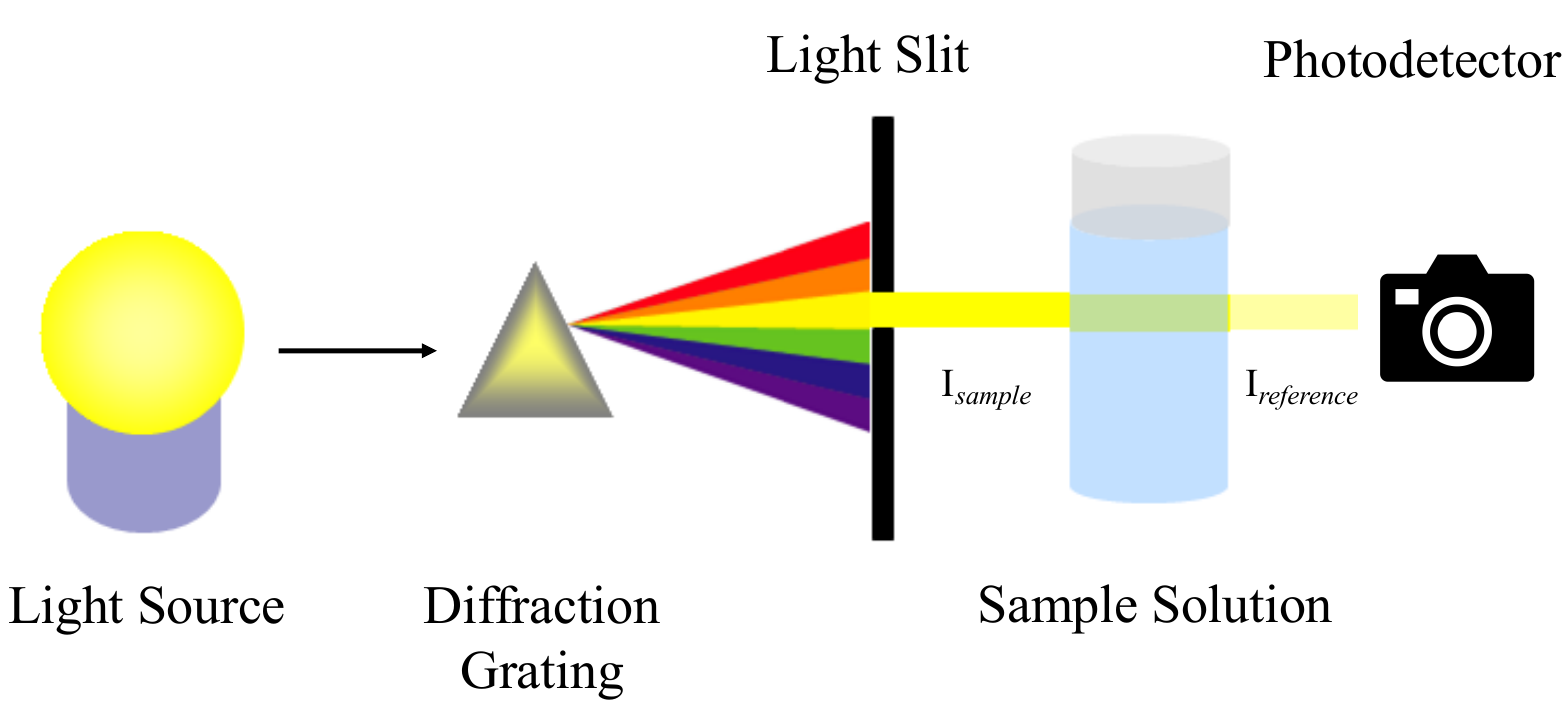}
    \caption{Basic spectrophotometer components, adapted from \cite{vo-215-2013}.}
    \label{fig:example-basic-spectro}
\end{figure}

Spectrophotometry is a measure of how much electromagnetic radiation, or light, a medium absorbs or transmits. A spectrophotometer is a device that measures the amount of photons (light intensity) absorbed and transmitted after it passes through a sample \cite{whatIsSpectro}. 
The interaction of a compound with electromagnetic radiation reveals its intrinsic properties and allows for identification and quantification. 
In our work, we utilize visible spectrophotometry, as a smartphone camera can capture the visible range of wavelengths.

A visible spectrophotometer typically contains a light source, a diffraction grating, a light slit, a sample holder, and a photodetector \cite{whatIsSpectro}. An example is shown in Fig. \ref{fig:example-basic-spectro}.
The \textit{light source} is a source of optical radiation that is either monochromatic and emits a singular color (wavelength) of light or a broadband source that evenly emits radiation across a range of wavelengths. A \textit{photodetector} captures the light after it passes (transmits) through the sample.
The photodetector measures light separated into individual wavelengths to determine light transmittance at each particular wavelength. Thus, either the light source is monochromatic, or there is a \textit{diffraction grating} to separate broadband light into all of its constituent wavelengths. A \textit{light slit} is a narrow, parallel opening for light to pass through.
The light slit can either act as a wavelength selector over diffracted light if placed after the diffraction grating (as in Fig. \ref{fig:example-basic-spectro}) or as a way to limit the intensity of the incoming light and the light scattering if placed before. 
The \textit{sample holder} holds the sample in place and is typically a clear cuvette that allows light to transmit through.

A spectrophotometer's photodetector measures the transmittance of a sample.
The transmission ($T$) of light by a compound at some wavelength is defined as follows: 
\begin{equation}
\label{eq:transmittance}
    T = \frac{I_{sample}}{I_{reference}},
\end{equation}
where $I$ is the intensity of light at that wavelength, as measured by the photodetector \cite{vo-215-2013}. Absorbance ($A$) is a measure of the light absorbed by a sample at a particular wavelength. It is calculated using the obtained transmittance \cite{vo-215-2013}: 
\begin{equation}
    A = -log_{10}(T) = -log_{10}(\frac{I_{sample}}{I_{reference}})
\end{equation}

The absorbance spectra of a compound in the visible range 
can be used to determine its identity or concentration. The latter can be done via the Beer-Lambert Law (Beer's Law), which states that there is a linear relationship between the absorbance and concentration of a sample. Thus, if there is a linear relationship, then Beer's Law is defined as the following:
\begin{equation} \label{eq:beer}
    A = \epsilon l c,
\end{equation}

where $A$ is still absorbance, $\epsilon$ is the molar absorptivity, $l$ is the path length, and $c$ is the concentration \cite{vo-215-2013}. Related work utilizes the term spectroscopy to refer to a spectrophotometer excluding a light source, or they are sometimes used interchangeably.

\subsection{Smartphone-based Spectrophotometry}
There is a growing demand for miniaturizing spectrophotometry to bring testing directly into the home, clinic, and field. A review on portable spectroscopy by Crocombe \cite{crocombe2018} highlights the influx of portable and handheld spectrometers, describing a wide berth of applications ranging from point-of-care devices in healthcare, law enforcement, and field analysis. In response to this trend, manufacturers have adapted laboratory-grade spectrophotometers for portability (Hach) \cite{hach_dr900}, (Ocean Optics) \cite{ocean_optics}, (HunterLab) \cite{hunterlab}. Mobile spectrophotometry is being integrated into wearable devices to further integrate the sensing modality into everyday life \cite{watson_lumos, preejith2016}. Several U.S. patents have advanced mobile spectroscopy, with some specifically incorporating smartphone integration \cite{hruska2019patent, wang2008patent, blair2021patent, zhang2010patent, hannel2011patent, blair2016patent}.



To make spectrophotometry more accessible, smartphones are increasingly utilized in designs, with researchers taking advantage of their high-resolution camera, light-emitting-diode (LED) flash, and processing power. 
However, we find that existing designs are still complex, costly, and require inaccessible components. Although there are DIY smartphone-based spectrophotometer designs for educational purposes \cite{bruininks-inexpensive-2022, Hosker2018, Bogucki2019, Jarujareet2023, Grasse2016}, they are generally concerned with making simple, cost-effective devices that can be replicated by students. As a result, many do not verify their devices' accuracy with benchtop laboratory-grade spectrophotometers, limiting their potential use in applications that require high precision.

Additionally, most smartphone-based spectrophotometer applications are used in conjunction with calorimetry, which requires a carefully prepared chemical reaction involving particular reagents to quantify and detect the compound of interest.
In an application of food chemistry, smartphone-based spectrophotometers have been designed to analyze colorimetric assays for detecting vitamin C (ascorbic acid) \cite{AGUIRRE2019, Kong2020}. 
This approach requires the addition of a chemical reagent that reacts with ascorbic acid present in the sample to change color. Aguirre et al. \cite{AGUIRRE2019IBU} took a similar approach to quantify the anti-inflammatory drug Ibuprofen for pharmaceutical applications. Ding et al. developed a smartphone-based spectrophotometer with an optical fiber bundle, demonstrating its ability to quantitatively measure creatine levels \cite{Ding2018}. A different group designed and tested a smartphone-based spectrophotometer for calorimetric analysis of glucose and human cardiac troponin I, integrating the phone's camera and flashlight \cite{Wang2016calor}. 
Silva et al. used GoSpectro \cite{gospectro}, a commercially available spectrometer attachment for smartphones, to design a smartphone-based spectrophotometer to estimate copper and iron concentrations in water for water pollution applications \cite{Silva2022}. Yu et al. were the first to use smartphone-based spectrophotometry for reading out fluorescence-based biological assays \cite{Yu2014}. Wang et al. developed a smartphone optosensing platform using a DVD grating to detect neurotoxins \cite{Wang2016}. They claim to be the first to use a DVD (digital versatile disc) grating as the diffraction grating proponent and compare its performance with that of a commercial-level grating.
A review on smartphone spectrometers lists several other applications \cite{Mcgonigle2018}.




Overall, there is an opportunity to advance the utility of accessible, smartphone-based spectrophotometry for nutrition assessment.
Our prototypes explore the potential of smartphone-based spectrophotometry to detect and quantify nutrients directly in solution, an application that has been ignored in favor of more complex colorimetric methods.
This is a critical step towards the most accessible manner of assessing nutrients within the human body.
We thoroughly evaluate our prototype and gather insights for future designs by considering various light sources and reference spectra calculation methods, which no prior work has formally presented.





\section{Prototype Development}
\subsection{First Generation Prototype With a Compact Fluorescent Bulb Light Source}

\subsubsection{Design} The first generation of our prototype is a minimum viable design to test feasibility. 
Because smartphone-based spectrophotometry has not been applied to analyze nutrients (without the help of an assay or colorimetric reagents), we wished to use this prototype as a simple proof-of-concept to see if nutrients could be analyzed and which ones would make the best targets. 

\begin{figure}
    \centering
    \includegraphics[width=0.7\linewidth]{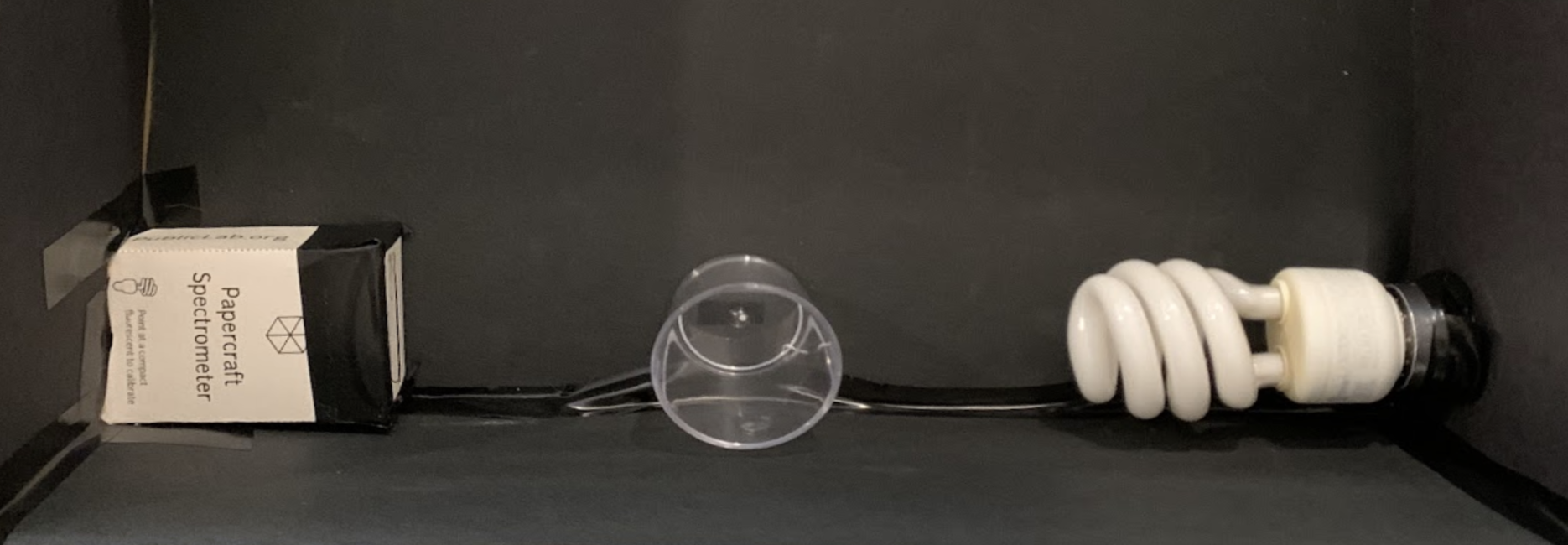}
    \caption{First generation device prototype seen from above. The constructed spectrophotometer is on the left, and a CFL source is on the right. It is lined with black construction paper and covered with a lid during operation to limit outside light. The sample is placed in the middle.}
    \label{fig:gen1}
\end{figure}

The design consists of a light source, which emits light through the sample and into a light slit. The incoming light is then separated into individual wavelengths by the diffraction grating and imaged with the smartphone camera. The design uses at-home materials and is inserted into a blacked-out box to minimize outside light (Fig. \ref{fig:gen1}). The slit and diffraction grating were implemented in an assembly from PublicLab \cite{jeffrey_warren_build_2024}. The slit was constructed using two razor blades with their sharp, parallel edges fastened a few millimeters apart. The diffraction grating consists of the polycarbonate substrate of a DVD-R. This assembly was inserted into one end of the box, opposite of a compact fluorescent (CFL) bulb. A sample in a clear plastic container was placed in the middle, in the path of light.
When a smartphone is laid against the diffraction grating on the outside of the light-proof container, it can capture an image of the light spectrum transmitted through the sample. 

\subsubsection{Signal Processing and Analysis} 
\begin{figure}
    \centering
    \includegraphics[width=0.75\linewidth]{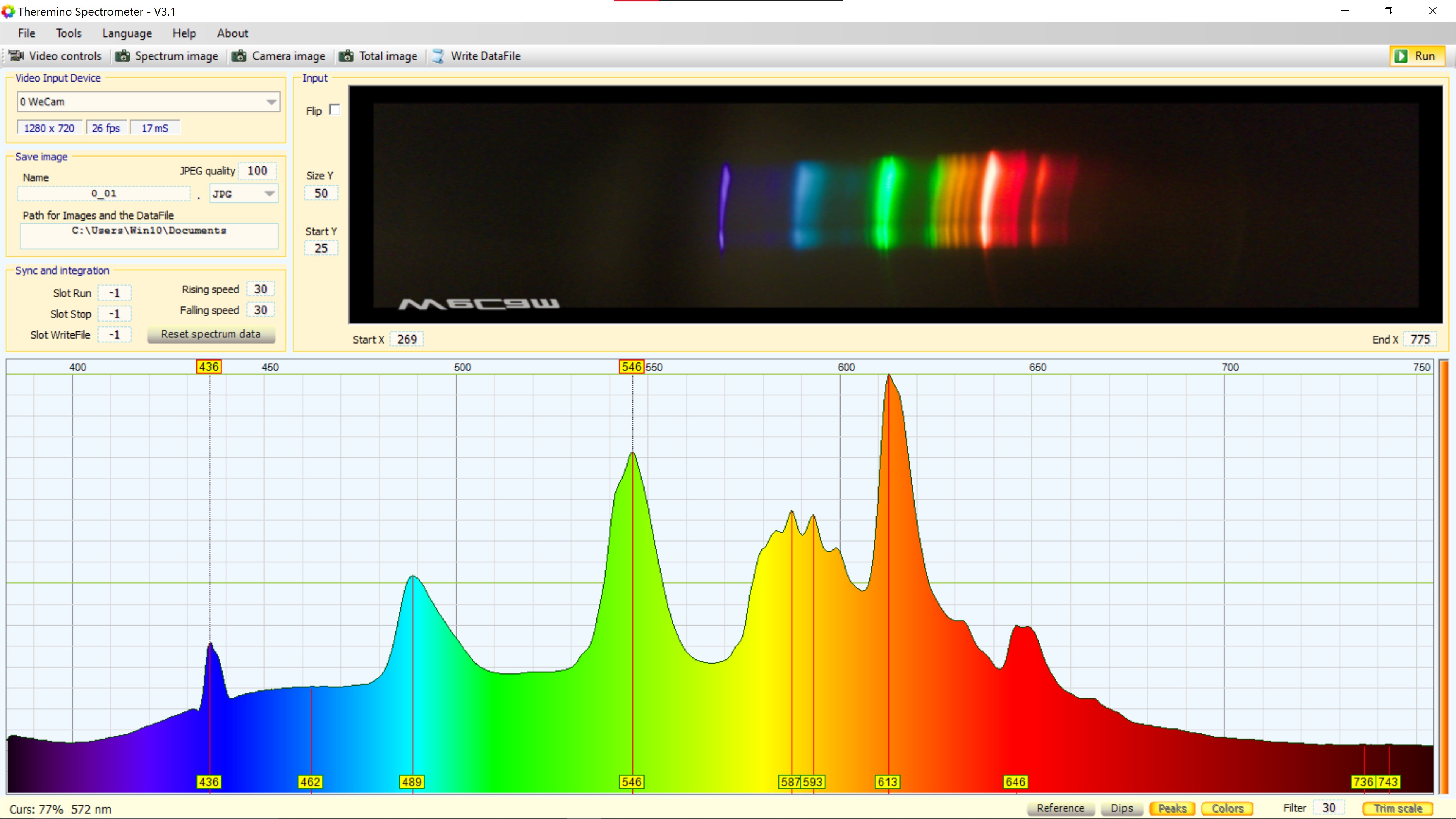}
    \caption{Blank sample (distilled water) as imaged with the first generation prototype and analyzed by the Theremino Spectrometer software. CFL bulb peaks labeled for calibration.}
    \label{fig:Theremino_calibration}
\end{figure}

We used the open-source Theremino Spectrometer software to extract a sample's transmittance spectra (Eq. \ref{eq:transmittance}) from a smartphone image \cite{noauthor_theremino_nodate}. 
Utilizing Theremino requires manual steps. The user first connects to a camera and performs the calibration step by selecting a region of interest (ROI) from the spectra image and labels known emission peaks (Fig. \ref{fig:Theremino_calibration}). A CFL bulb's consistent emission peaks allowed us to calibrate our device by mapping the location of each peak in horizontal pixels of the image to the wavelength at which it is known to appear. Once calibrated, a blank sample is placed in the spectrophotometer to obtain $I_{reference}$ from Eq. \ref{eq:transmittance}. Then, the user can insert the test sample, collect data for $I_{sample}$, and calculate the transmittance and absorbance spectra.


\subsubsection{Feasibility Evaluation}
We conducted a feasibility study to evaluate the first-generation smartphone-based spectrophotometer and address \textit{RQ1}. 
The experiment selected a set of three nutrient supplements: vitamins B2, B7, and B12.
If these nutrients could be differentiated using a simple proof-of-concept, a more refined design capable of quantification would be worth perusing.
We dissolved each supplement into 40 mL of distilled water and filtered the solution to remove insoluble material from the supplement formulation. We used the setup depicted in Figure \ref{fig:gen1} to evaluate the samples. A portion of the sample volume was transferred into a clear plastic container and placed in the optical path of a CFL bulb. The smartphone (iPhone XR) was held steady and even against the diffraction grating, and a picture of the sample's spectra was taken. This image was analyzed with the Theremino Spectrophotometer and compared to a blank sample (distilled water) to obtain the absorbance spectra. These steps were repeated for each of the three nutrient supplements.

\subsubsection{Observations}
\begin{figure}
    \centering
    \includegraphics[width=0.8\linewidth]{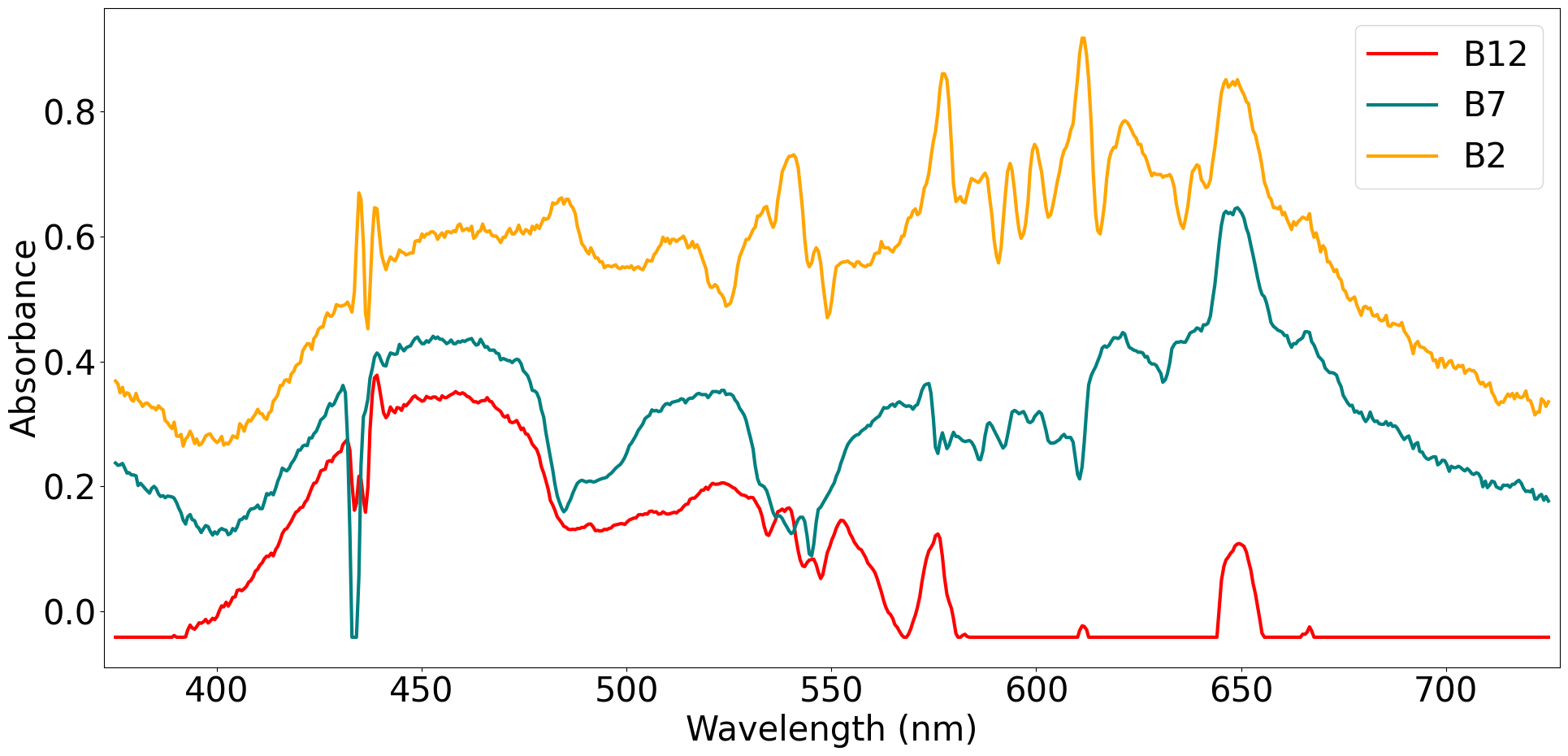}
    \caption{The three test nutrients had distinct absorbance spectra in the first-generation prototype. Results for vitamin B12 appeared to be the most similar to the physical properties of the solution (a pink-red color). Prevalent noise in the spectra necessitated further testing and an improved design.}
    \label{fig:gen1-feasability}
\end{figure}

Our results demonstrate that each vitamin has distinct absorbance spectra, likely representing their physical color (Fig. \ref{fig:gen1-feasability}). Vitamin B12 is the best example of this phenomenon, exhibiting lower absorption in longer wavelengths of light (past $\sim$582 nm). This observation is consistent with the physical color of the vitamin B12 sample, which was pinkish-red. Vitamin B7, which was white and more opaque in appearance, exhibited greater absorbance across the visible spectrum. This aligns with our expectation that higher opacity is associated with higher absorbance. The results for vitamin B2 were more difficult to interpret. While the solution had a distinct and quite opaque yellow color, a corresponding, dominant transmission peak in the orange-yellow wavelengths is not evident in the spectrum. However, the high absorption across the visible spectra is consistent with its opacity. These findings may be due to a combination of having too large a concentration of B2, as well as the inherent design limitations of this first-generation prototype.

\subsubsection{Implications} This first-generation prototype demonstrated the potential of smartphone-based spectrophotometry to differentiate between the test set of nutrients (\textit{RQ1}). However, the noise and imprecision should be noted. The following limitations of the prototype could be addressed to fully realize this potential: 1) the bulky, non-portable design, 2) the error induced by the need to manually position the smartphone and cuvette for each image, and 3) the slow process of analyzing each image with Theremino Spectrophotometer. We also determined that vitamin B12 would be an optimal candidate for further testing because our spectrophotometer most accurately characterized its distinct pink-red appearance. In addition, B12 is readily dissolved in water, which facilitates experimentation more easily. 


\subsection{Second Generation Prototype with Portable Design and Semi-Automatic Signal Processing}

\subsubsection{Design} 
\begin{figure}
    \centering
    \includegraphics[width=1\linewidth]{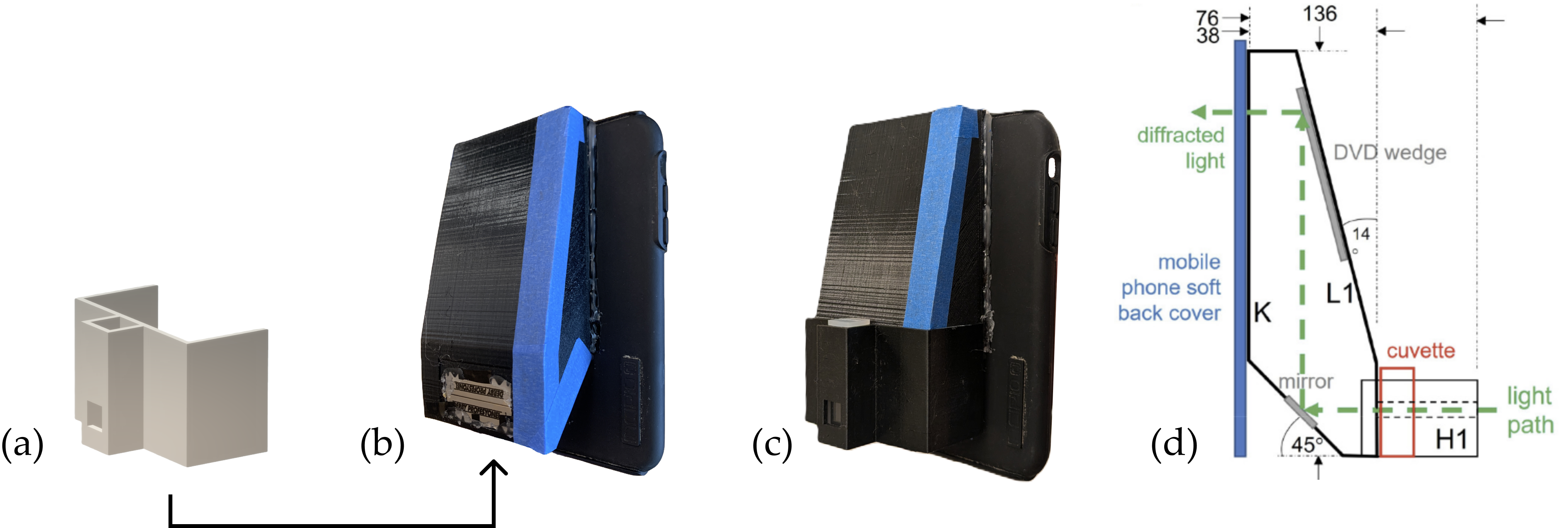}
    \caption{Replicated spectrophotometer assembly where (a) the custom-designed cuvette holder attaches to (b) the 3D printed design, making up (c) the final assembly. The (d) design schematic is reprinted from the source \cite{bruininks-inexpensive-2022}.
    }
    \label{fig:gen2}
\end{figure}

To address the first two limitations posed by our initial prototype, we implemented a design by Bruininks et al. \cite{bruininks-inexpensive-2022}. The design itself consists of a 3D-printed, periscope-like assembly (Fig. \ref{fig:gen2}). Again, two razor blades are used to create a narrow, straight slit for light to pass through. Internally, the light from the slit bounces off a mirror and into the diffraction grating, for which we continue to use the polycarbonate substrate of a DVD-R. The incident (reflected) light from the diffraction grating is directed into the smartphone camera sensor (Fig. \ref{fig:gen2}d). The housing is attached directly onto a smartphone case such that the smartphone’s camera is positioned in the center of the spectrometer’s opening. This integration addresses limitations 1 and 2 from the first generation prototype: it is more compact and portable, and it eliminates errors arising from the need to reposition the smartphone. 


The original design by Bruinunks et al. consists of an additional cuvette and LED holder accessory \cite{bruininks-inexpensive-2022}. However, we found it requires steady positioning on a flat surface with the smartphone held flush to the opening, which made it challenging to reproduce results. Additionally, we wanted a design that would easily allow for various light source experimentation. Therefore, to further improve precision in setting up the sample for analysis, we developed a custom, 3D-printed cuvette holder that attached directly to the main assembly
(Fig. \ref{fig:gen2}b). To accommodate larger, more intense light sources (e.g. a LED bulb) and allow light to easily pass through, two relatively large, square openings were integrated on either side of the cuvette. One faced the light source and the other faced into the device.


\subsubsection{Signal Processing and Analysis}
\begin{figure*}
    \centering
    \includegraphics[width=\textwidth]{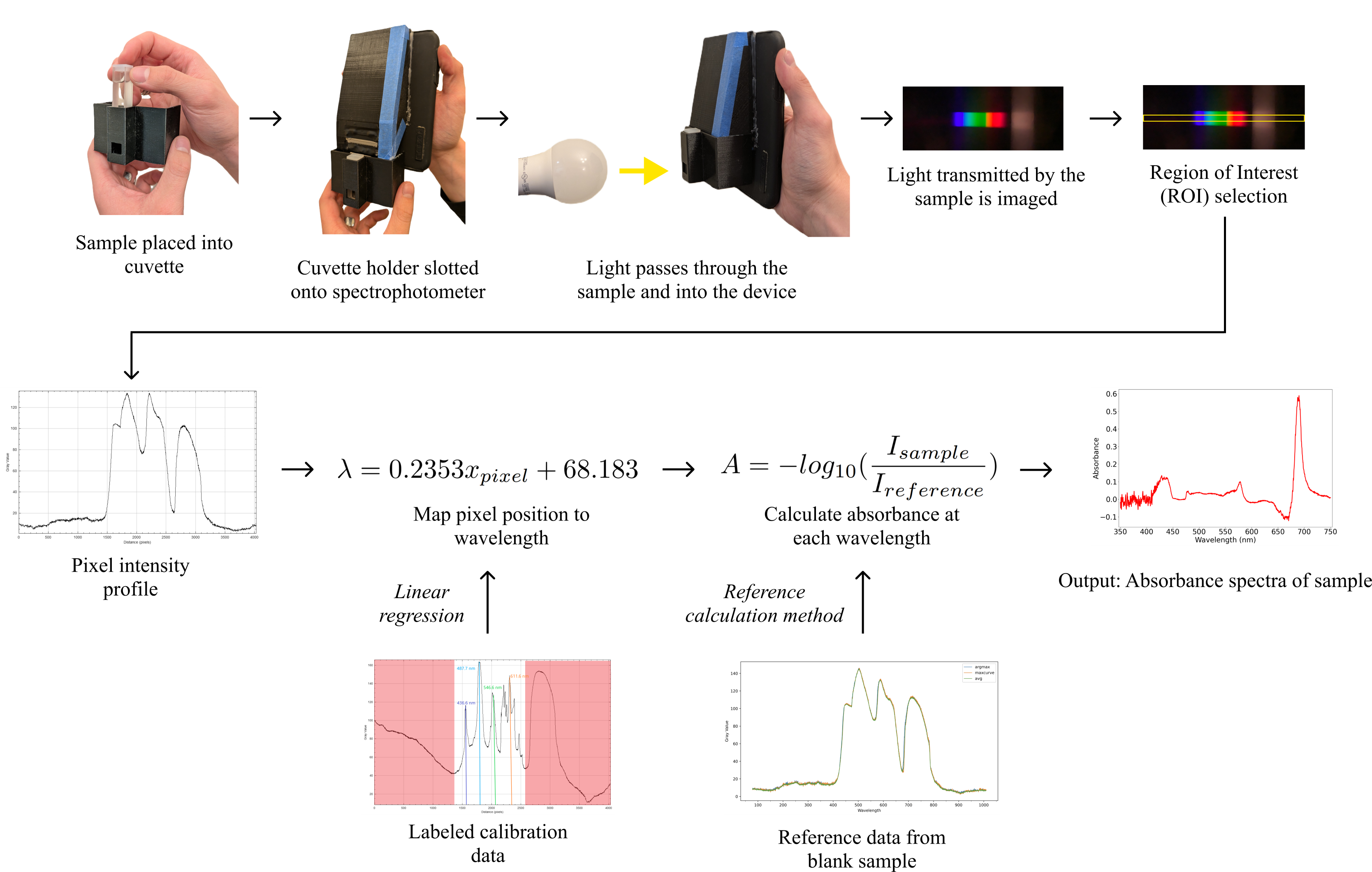}
    \caption{The processing and analysis pipeline for generating absorbance spectra of a sample from a smartphone captured image.}
    \label{fig:pipeline}
\end{figure*}

For this prototype to be more viable as a mobile tool and to facilitate experimentation at a greater scale, a more streamlined process of extracting quantitative information from the obtained smartphone images is necessary. Thus, to address the third limitation of the first-generation prototype, we implemented our own semi-automatic, Python-based signal analysis pipeline (Fig. \ref{fig:pipeline}).

The fundamentals of the process are similar to the previous approach, but executed on a lower level. Given a horizontal image of the diffracted light, a consistent ROI is selected. This is the integration region over which the intensity profile is determined (a function of the average y-axis gray value over the pixels in the x-axis, where the gray value is a measure of intensity). The software ImageJ was used to calculate this profile. Using a calibration image of a CFL bulb and a blank sample, we fit a linear equation to labeled pixel positions of the known emission peaks of a CFL (436.6, 487.7, 546.5, and 611.6 nm). The result is a linear equation that maps the x-axis location of pixels in an image to wavelengths. Our purpose-built pipeline provides the ability to analyze hundreds of samples at scale, and to analyze and tune several parameters involved in the signal analysis process.

\section{Experimental Evaluation}


\subsection{Experimental Setup}
\subsubsection{Nutrient Preparation} To ensure the robustness of our experiments, we tested two brands of vitamin B12 supplements: Trader Joe's (TJ) and Walgreen's (WG). Multiple concentrations of vitamin B12 were prepared for each brand as follows:
\begin{itemize}
    \item Supplements were dosed, ground, and dissolved into distilled water at a concentration of 20 $\mu$g/mL, creating stock solutions.
    \item The stock solutions were filtered through a paper filter to remove any insoluble material that remained from the supplement.
    \item The filtered stock solutions were diluted to four lower concentrations (10, 4, 2, and 1 $\mu$g/mL).
\end{itemize}

\subsubsection{Light Source} An impactful independent variable for a spectrophotometer is the light source. An ideal light source emits light radiation at a uniformly strong intensity over the visible spectrum. Considering device accessibility, we evaluated four consumer-available light sources: a CFL bulb, an LED bulb, a single LED diode, and an LED diode matrix. LEDs are well represented by this sample because of their ability to generate light in a more even intensity. 

\subsubsection{Reference Spectra Calculation}
Next, we experimented with different ways of defining the reference spectra, $I_{reference}$. Since a smartphone-based spectrophotometer is highly susceptible to inter-trial variance, a reliable method of ``zeroing-out" the device is needed. This can be achieved by taking multiple images of the blank sample and combining information from each to inform the reference spectra. Three different strategies were employed to define the reference spectra of the blank sample: 1) averaging the intensity across each blank sample, $I_{average} = \{ I_{\lambda}: \frac{1}{n} \sum_{i=0}^{n} I_{\lambda, i}\}$ 2) selecting the spectra with the sum largest intensity (maxcurve), $I_{maxcurve} = \arg\max_{I_{i}} \sum^{\forall \lambda} I_{\lambda, i}$ and 3) taking the maximum of the three intensities at each wavelength, $I_{max} = \{ I_{\lambda}: \max I_{\lambda, i}\}$.

\subsection{Experimental Procedure} We tested our original prototype and the second generation prototype utilizing our semi-automatic, Python-based signal analysis pipeline to extract the absorbance spectra data. In both cases, the same smartphone captured the images of the diffracted light transmitted through the sample (iPhone XR), and all samples were analyzed in a quartz cuvette designed for UV-visible spectrophotometry. All four light sources and the three strategies of reference spectra calculation were implemented and compared for both the first- and second-generation devices. 

Three images of each sample were captured using the spectrophotometer setup. Each image was taken immediately after the previous one, and their intensity profiles were then averaged. Additionally, three images of the blank sample (distilled water) were captured under different light source or prototype generation conditions. These images were combined with different reference spectra calculation methods.
The final dataset comprised 240 test sample images and 24 blank images. After the signal analysis pipeline was executed and the different reference spectra calculations were applied, 240 unique absorbance spectra were extracted for evaluation.


\subsection{Evaluation Criteria}
Qualitatively, ideal results would produce absorbance spectra with relatively low noise, well-defined absorbance peaks or valleys, high inter-trial reliability (low variance), and a clear ordering of concentrations. 
Quantitatively, the results should follow the Beer-Lambert Law, which demonstrates that absorbance at a given wavelength is directly associated with concentration (Eq. \ref{eq:beer}). 
This property is coupled with the clear ordering, as a result that follows the Beer-Lambert Law will have absorbance values that monotonically increase or decrease with respect to the sample concentration at a distinct peak or valley. 

The differences between various design parameters (\textit{RQ2}) and prototype devices are determined via visual inspection and qualitative analysis. Additional quantitative analysis is presented to formalize the performance of the best design against the laboratory spectrophotometer (\textit{RQ3}).

\section{Results}



\subsection{Effect of Different Light Sources}
By comparing absorption spectra within the same concentration, brand, and prototype conditions, we found that the single LED and LED matrix light sources performed poorly. These sources showed high inter-trial variability in most cases. Any peaks or dips that could be determined were of low quality, with concentrations out of order and/or at very similar absorbance values. The exception to this was the LED matrix samples, likely because of the methodological decision to select the single clearest diffraction pattern as the ROI, out of the 2-3 visible from the matrix of LEDs. 

\begin{figure}
    \centering
    \includegraphics[width=\linewidth]{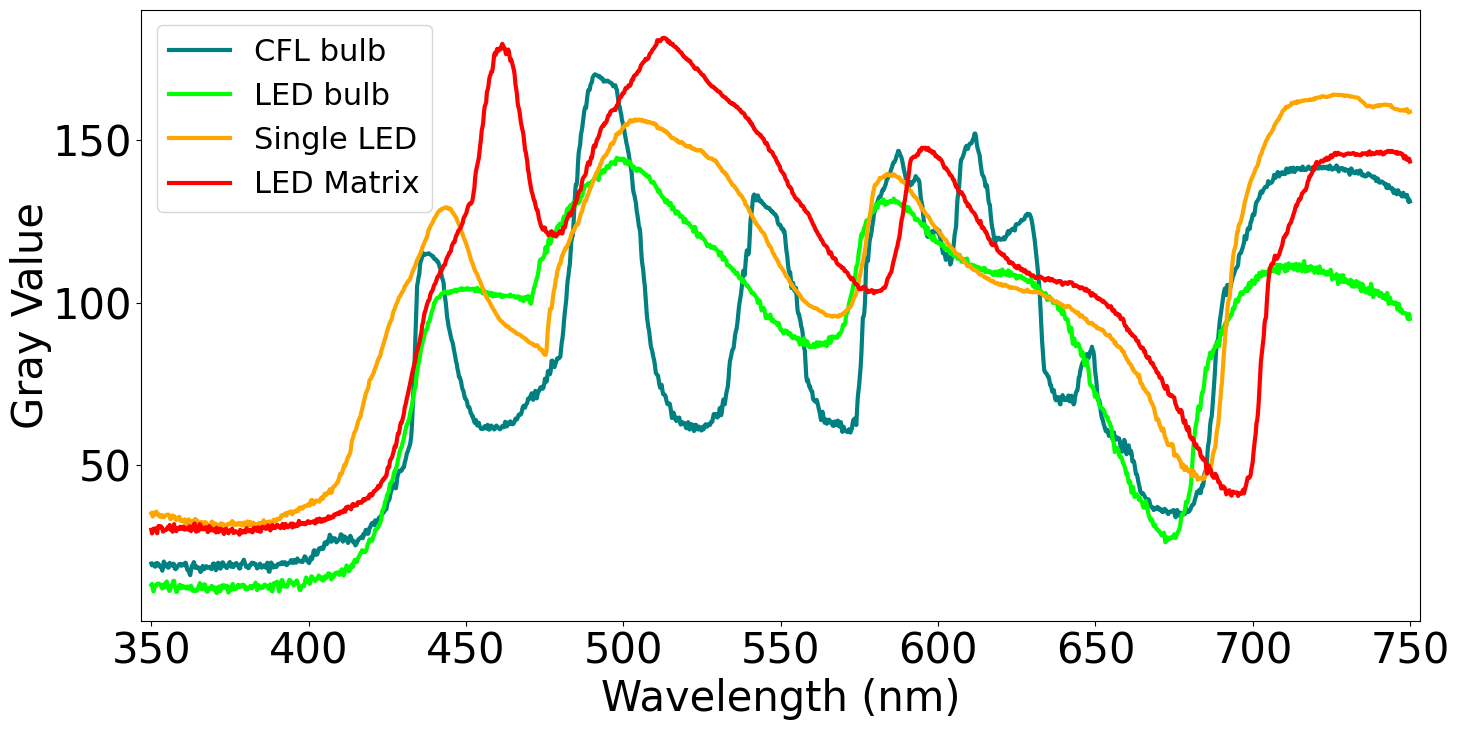}
    \caption{Reference spectra of each light source evaluated, as determined by the second generation prototype.
    }
    \label{fig:lights}
\end{figure}

The CFL and LED bulbs produced clear peaks and minimal noise when evaluated on test samples. The LED bulb showed a more uniform emission spectra, with less emission peaks and valleys, when evaluated on a blank (Fig. \ref{fig:lights}). The bulbs likely performed best because of the evenness of the light they emit, whereas the individual LEDs are more concentrated with a narrow degree of illumination. Therefore, an LED bulb is the most ideal light source for our device out of those tested. A detailed evaluation between the light sources is further described when comparing the second-generation device to the original, in Section \ref{comparison to first}.

\subsection{Effect of Reference Spectra Calculation}
\begin{figure}
    \centering
    \begin{subfigure}{0.48\columnwidth} 
        \centering
        \includegraphics[width=\linewidth]{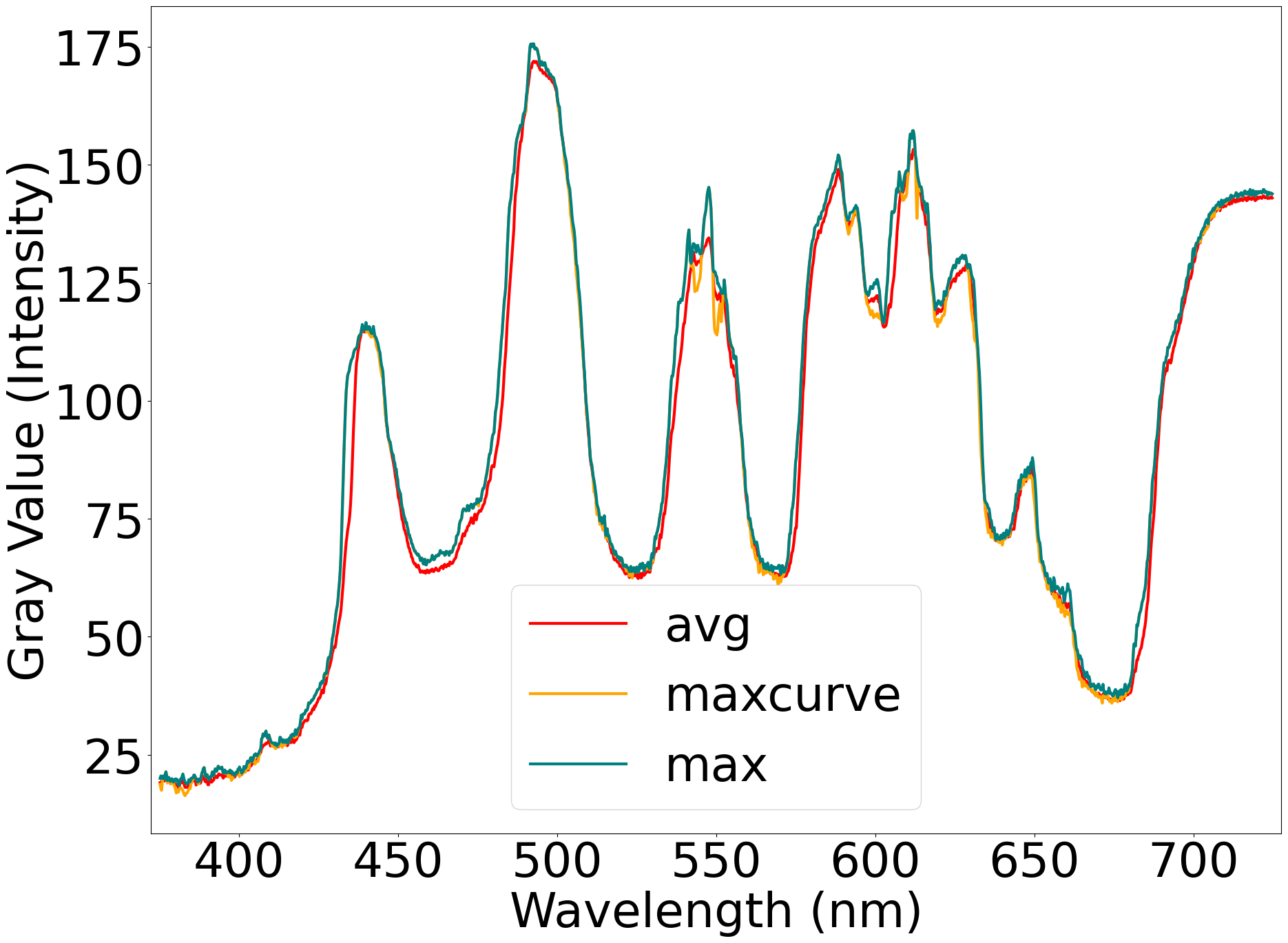}
        \caption{}
        \label{fig:ref_cfl}
    \end{subfigure}
    \hfill
    \begin{subfigure}{0.48\columnwidth} 
        \centering
        \includegraphics[width=\linewidth]{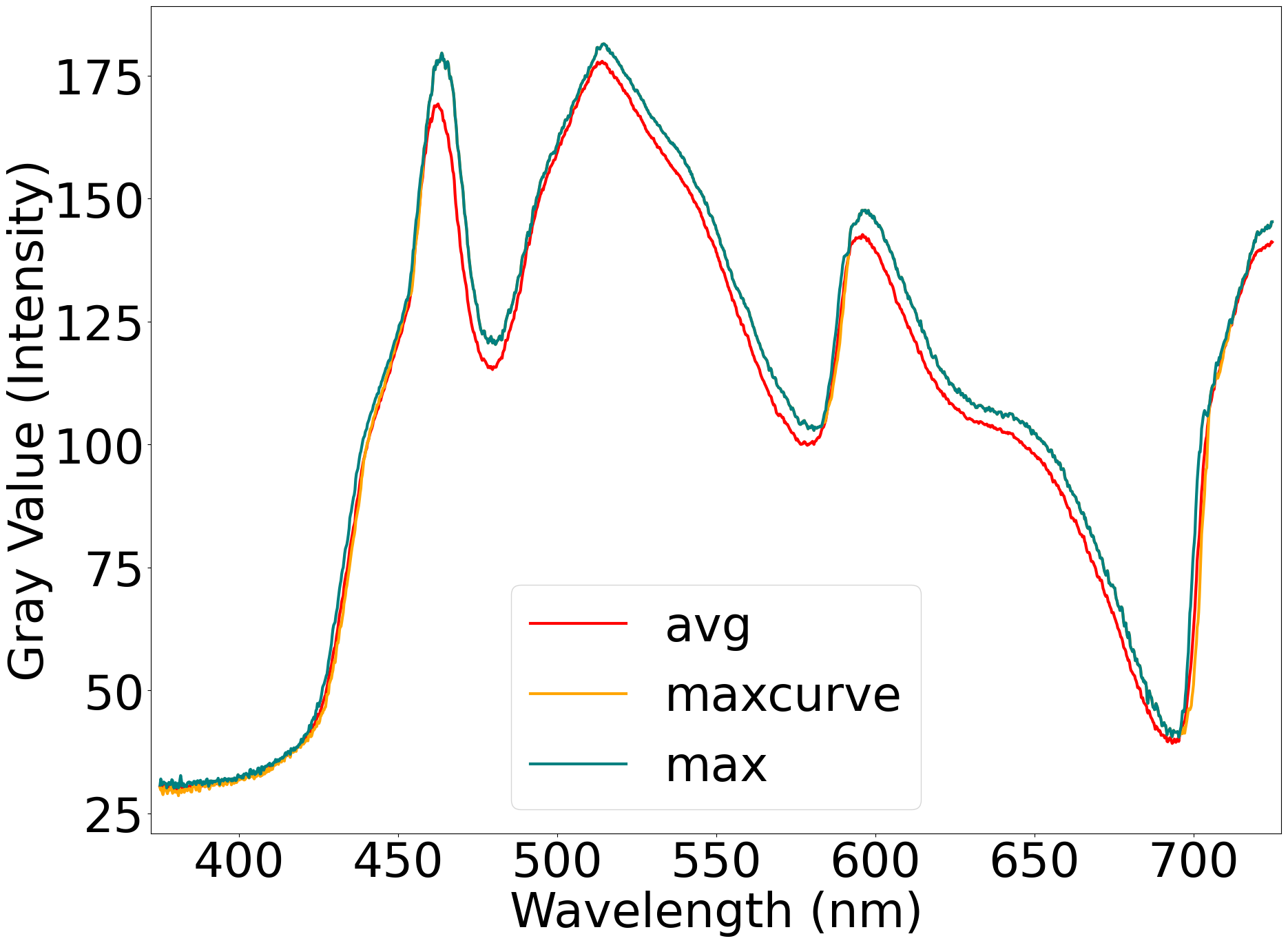} 
        \caption{}
        \label{fig:ref_worst_case}
    \end{subfigure}
    \caption{A comparison of the intensity profiles for the three reference spectra calculation methods. (a) The CFL bulb, with its many peaks, provides a representative example of the similarities between the proposed methods. (b) The LED matrix light source shows a worst-case scenario, with differences in peak amplitude but not location.}
    \label{fig:ref_comp}
\end{figure}

Reference spectra calculation methods were compared by plotting each of their intensity spectra together and auditing the peaks, valleys, and general shape (Fig. \ref{fig:ref_comp}). In each case, differences in the spectra of the reference 
were minimal. Most importantly, the shape of the spectra (peak and valley locations) are nearly identical for all methods (Fig. \ref{fig:ref_cfl}). More differences are seen in the intensity (gray values) of the spectra. As expected, $I_{max}$ results in the highest intensity and is often made up of the $I_{maxcurve}$ sample (Fig. \ref{fig:ref_worst_case}). The $I_{average}$ reference is generally lower in intensity than the other two, but not by much. 
At no point was there a difference in peak/valley location or the order of the spectra at that peak, only the amplitude (absorbance) was periodically affected. Therefore, we found that no particular approach was distinct when comparing the various reference spectra strategies.

\begin{figure*}
    \centering
    \includegraphics[width=1\textwidth]{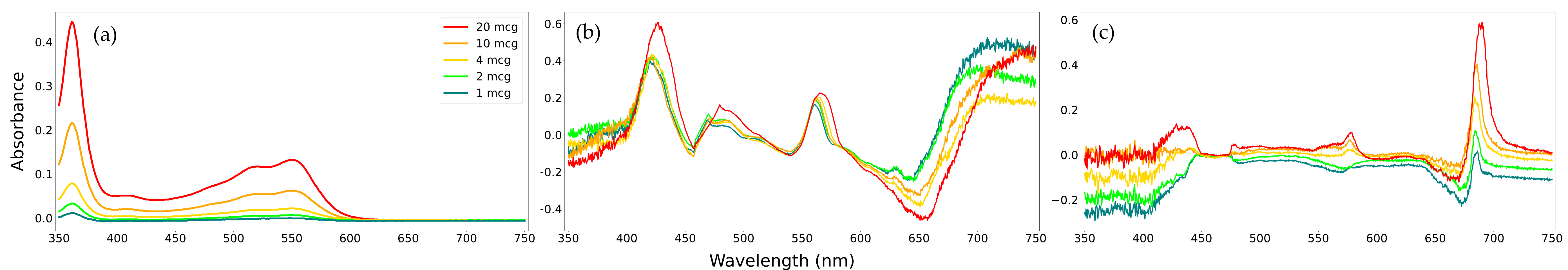}
    \caption{Absorbance peaks in the visible spectrum for different concentrations of vitamin B12 (WG) as determined by different devices. While the level of noise varies, the three devices show agreement on a peak within 550 to 575 nm. 
    (a) Laboratory device (ground truth) exhibits peaks at at 361 and 551 nm. 
    (b) The first-generation prototype (LED bulb) exhibits peaks at 422, 470, and 563 nm.
    (c) The second-generation prototype (LED bulb) exhibits peaks at 577 and 690 nm. Concentrations are $\mu$g/mL.}
    \label{fig:device-comp}
\end{figure*}

\subsection{Comparison to First Generation Prototype}\label{comparison to first}

\begin{table*}[t]\centering
\caption{Qualitative analysis of both prototype devices across different light sources and supplement brands. ``Outlying behavior" indicates when the absorbance spectra of a sample varies significantly from the others. $C$ refers to concentration and $A$ absorbance.}\label{tab:gen_comparison}
\scriptsize
\begin{tabular}{p{0.33in}|p{1.6in}p{1.4in}p{1.6in}p{1.33in}}
\toprule
&CFL &LED Bulb &Single LED &LED Matrix \\\midrule
Gen 1 
&Common peak at $\sim$580 nm (decreases with $C$); high variance; valleys red shift as $C$ increases \newline \textbf{TJ:} Outlying behavior in 1 and 10 $\mu$g; 4 and 20 $\mu$g have similar peak $A$ \newline \textbf{WG:} Outlying behavior in 4 $\mu$g; 4 and 2 $\mu$g have similar peak $A$ 
&\textbf{TJ:} High variance; outlying behavior in 2 $\mu$g, no well-defined peaks or valleys \newline \textbf{WG:} Peak at $\sim$560 (increases with $C$); valley at $\sim$650 (decreases with $C$; 4 \& 10 $\mu$g out of order); red shift as $C$ increases 
&Most variance of all results \newline \textbf{TJ:} Peak at ~575 (decreases with $C$; 20 $\mu$g does not follow trend) \newline \textbf{WG:} Valley at ~625 (increases with $C$; 2 and 4 $\mu$g out of order)
&\textbf{TJ:} High variance, no peaks or valleys with good order \newline 
\textbf{WG:} Peak at ~600 (increases with $C$), valley at ~575 (decreases with $C$; 4 $\mu$g out of order for both); some outlying behavior in 10 $\mu$g trials
\\\midrule
Gen 2 
&\textbf{TJ:} Clear peak at $\sim$560 (increases with $C$); less separated peak at $\sim$505 (increases with $C$; clear order on right base at $\sim$515), peak at $\sim$570 (increases with $C$) \newline \textbf{WG:} Outlying behavior in 10 $\mu$g (consistently high $A$) and 20 $\mu$g (low $A$); no peaks or valleys with good order 
&Most level spectra (sample matches reference well) \newline \textbf{TJ:} Similar shape to WG, but no order; outlying behavior in 10 and 20 $\mu$g (lower $A$ than expected) \newline \textbf{WG:} Clear peaks at $\sim$575 (increases with $C$) and $\sim$690 (increases with $C$) 
&Fairly level spectra; outlying behavior in 10 $\mu$g trials (higher $A$ than expected) \newline
\textbf{TJ:} Valley at ~420 (increases with $C$; 4 and 2 $\mu$g have close $A$), peak at ~675 (increases with $C$; 20 and 10 $\mu$g out of order) \newline
\textbf{WG:} Outlying behavior in 1 $\mu$g (higher $A$ than expected)
&Highest variance in Gen 2 trials \newline
\textbf{TJ:} No peaks or valleys with valid order \newline
\textbf{WG:} Valley at ~690 (decreases with $C$; \textit{interestingly, this is a peak for 1 $\mu$g and avg reference calculation})
\\
\bottomrule
\end{tabular}
\end{table*}

We characterize the differences between results for the two prototypes across different light sources (Table \ref{tab:gen_comparison}). Figure \ref{fig:device-comp} illustrates results from the first generation (Fig. \ref{fig:device-comp}b) and the second generation (Fig. \ref{fig:device-comp}c) with the LED bulb source. 

For our first-generation prototype, at least one peak between 560-600 nm can be observed with any light source. In the CFL bulb and LED matrix sources, this peak has decreasing concentration with amplitude, and for the LED matrix, the peak is only observable in the TJ samples. For the LED bulb 
and single LED sources, this peak has increasing concentration with amplitude and is only observable in the WG samples. LED-based sources in this generation shared a similar characteristic valley in the WG samples between 575 - 650 nm. In all cases with this light source, two concentration bands were out of order which always included the 4 $\mu$g band. With the LED matrix source, concentration increased with the absorption magnitude (depth) of the valley while it decreased with the others.

For our second-generation device, the CFL and LED bulb 
sources had limited similarities in observed peaks. Peak locations were at 560 and 575 nm, respectively, and concentration increased with peak amplitude. Concerningly, the CFL peak was only observed for TJ samples. The peak for the LED bulb at 575 nm was prominent for WG samples 
, but less so for TJ samples 
. The LED bulb 
and LED matrix sources also had a similar peak. This one was at 690 for the bulb and 675 for the matrix, again increasing in concentration. This peak was only observed for WG samples, but the matrix source had the 20 and 10 $\mu$g bands out of order. This error was commonly observed in the CFL bulb and single LED light sources as well. Since the 20 $\mu$g solution has a more vivid, pink-red color than the 10 $\mu$g, we expect its absorbance be greater, but this assumption is sometimes contradicted by the results (Table \ref{tab:gen_comparison}). This indicates that our device was better able to identify a nutrient than quantifying it.

Although both devices are subject to some degree of inter-trial variance, the second-generation prototype trials consistently showed less variance, especially when the optimal light source (LED bulb) was used. In addition, results from the second-generation device contain less noise induced by the light source than the first-generation. This is evident when comparing Figures \ref{fig:device-comp}b (first generation) and \ref{fig:device-comp}c (second generation). Both figures profile WG samples using identical light sources (LED bulb) and reference calculation methods (average), but the first-generation prototype exhibits two additional peaks at 422 and 477 nm.
The reference spectra for the LED bulb has peaks in similar locations 
to the first-generation device: 436 and 492 nm. Thus, we can infer that these inconsistent peaks result from the first-generation prototype capturing skewed images that manifest shifts in the absorbance spectra between trials.
As a result, the transmittance is not properly calculated (Eq. \ref{eq:transmittance}), and the spectra of the test sample are overpowered by the light source.
This quality is also evidenced by the apparent redshift (skew towards longer wavelengths) in the 20 $\mu$g/mL sample (Fig. \ref{fig:device-comp}b).
By contrast, the second-generation prototype can more reliably capture images of spectra, which is demonstrated by the lack of intense peaks caused by the light source (Fig. \ref{fig:device-comp}c).

\subsection{Comparison to Laboratory Spectrophotometer}
To validate the second-generation prototype and address \textit{RQ3}, we evaluated the performance of our smartphone-based spectrophotometer device against a laboratory UV-visible spectrophotometer (Hach DR6000). For this comparison, we select the most optimal design for the described smartphone-based spectrophotometer: LED bulb light source and averaged reference calculation.
For this set of experiments, we used a vitamin B12 supplement from Nature's Bounty (NB). 
The procedure was identical in all other aspects.

\begin{figure*}
    \centering
    \includegraphics[width=1\linewidth]{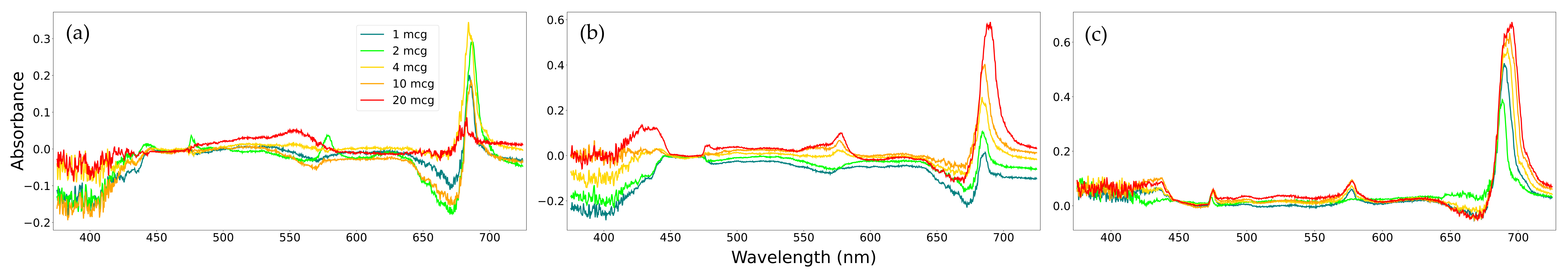}
    \caption{The second generation device is able to demonstrate an absorbance peak at $\sim$577 nm for all supplement brands, but the quality of this peak varies. (a) TJ brand supplements (b) WG brand supplements (c) NB brand supplements.}
    \label{fig:supplement-comp}
\end{figure*}

We first used the laboratory spectrophotometer to compare  WG and NB. The laboratory device revealed no obvious differences. Each had a prominent peak at 361 nm, just into the UV spectrum, and a flatter peak at 551 nm (see Fig. \ref{fig:device-comp}a for WG, which is representative). The inter-trial variance was negligible.
Comparing this to results from the smartphone device, we see that all three supplements (TJ, WG, and NB) have identical peak locations ($\sim$577 nm) and similarly-shaped absorbance spectra (Fig. \ref{fig:supplement-comp}). However, there is also considerable variance when it comes to the peak absorbance of each sample concentration. Noise is especially evident towards the boundaries of the visible spectrum ($<$450 and $>$650 nm), where the prototype device is unable to accurately measure transmittance (Fig. \ref{fig:device-comp}c).

\begin{figure}
    \centering
    \begin{subfigure}{0.49\columnwidth} 
        \centering
        \includegraphics[width=\linewidth]{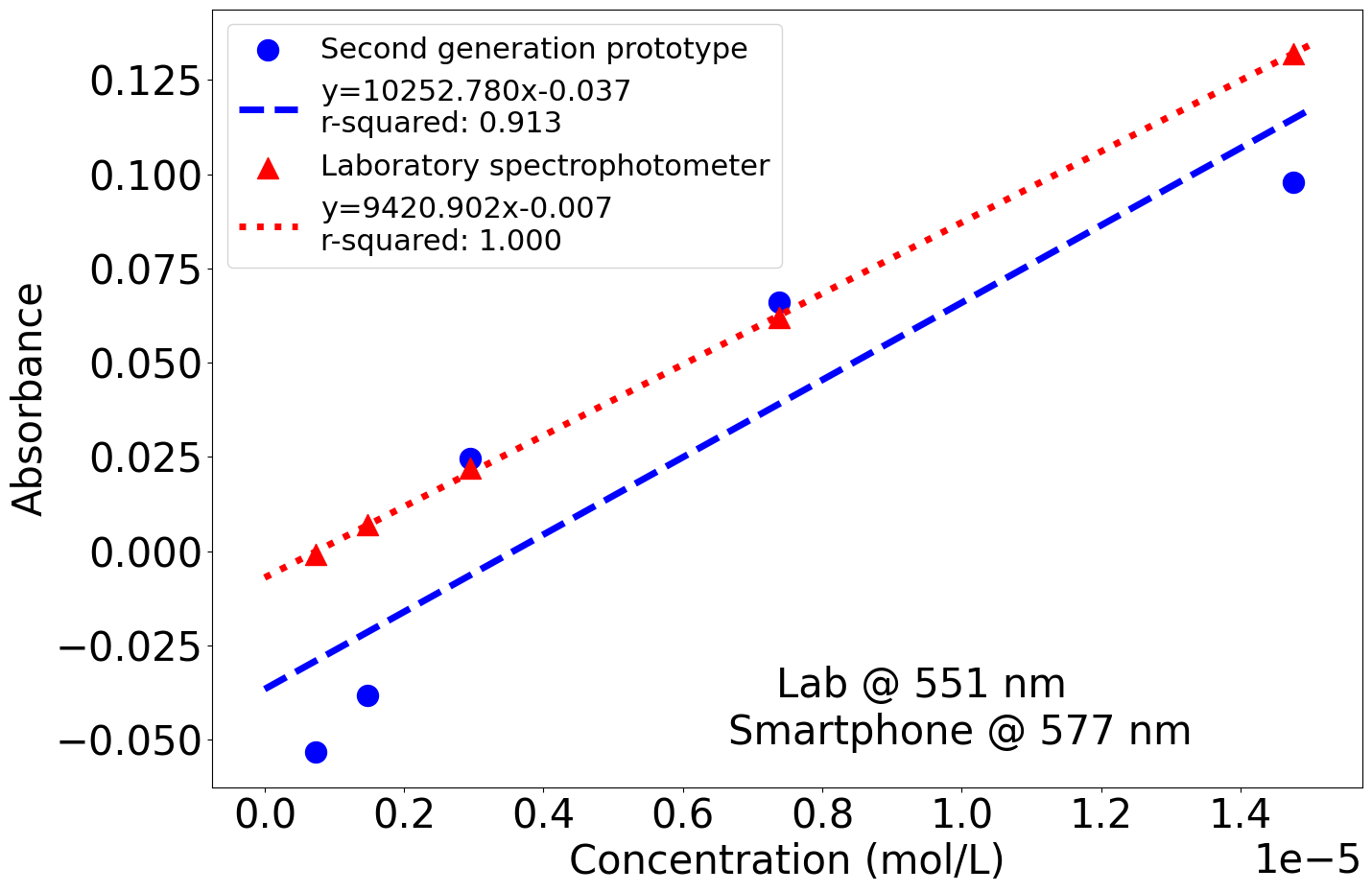}
        \caption{}
        \label{fig:beer_WG}
    \end{subfigure}
    \hfill
    \begin{subfigure}{0.49\columnwidth} 
        \centering
        \includegraphics[width=\linewidth]{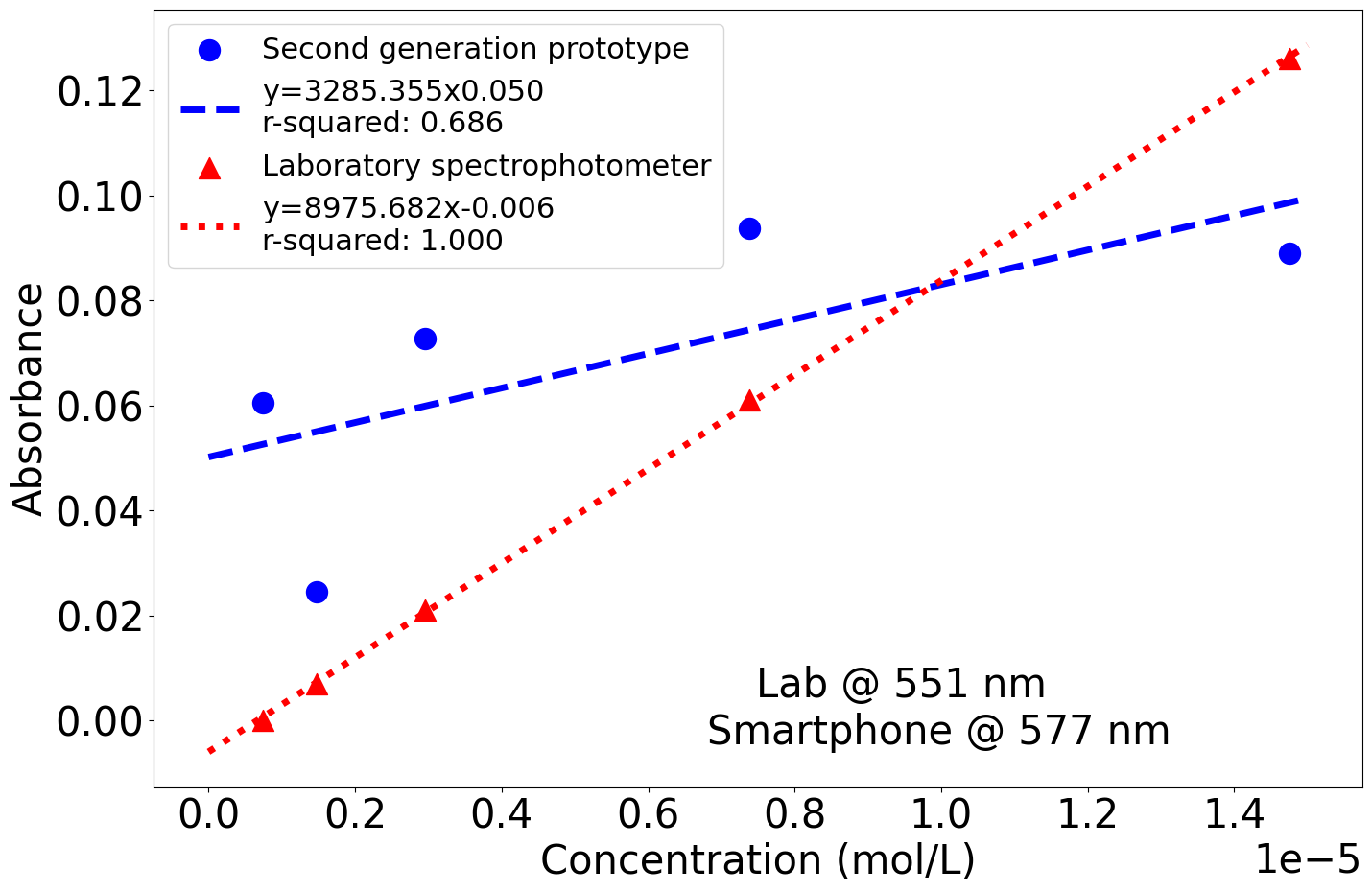} 
        \caption{}
        \label{fig:beer_NB}
    \end{subfigure}
    \caption{Relationship between absorbance and concentration for the laboratory spectrophotometer and the second generation prototype (LED light source, averaged reference, assessed at the most comparable absorbance peaks). (a) WG brand supplement (b) NB brand supplement}
    \label{fig:quant_comp}
\end{figure}

Importantly, however, the peak at $\sim$577 nm is only $\sim$20 nm red-shifted from the laboratory device results and demonstrates agreement with the laboratory device, as well as the first-generation prototype (for WG). Quantitative analysis confirms this for the WG samples, with regression analysis on the second generation smartphone device at 577 nm yielding an $R^2$ of 0.913 versus 1.0 for the laboratory device (Fig. \ref{fig:beer_WG}). The results from the second-generation prototype for the NB samples are less accurate, with an $R^2$ of 0.686 (1.0 for the laboratory device; Fig. \ref{fig:beer_NB}). This analysis means that spectrophotometric results from the present work follow the Lambert-Beer Law, with some degree of variance (Eq. \ref{eq:beer}). If provided with a vitamin B12 sample of unknown concentration, the prototype described herein can feasibly use the absorbance measured at 577 nm to determine the true concentration of the sample with an accuracy of 68.6\% - 91.3\%.

\section{Discussion}

Toward an accessible method of nutrient status assessment, we presented two early smartphone-based spectrophotometer prototypes for direct identification and measurement of nutrients. We experimented with various parameters and found that the LED bulb light source is more optimal and that all tested reference spectra strategies were similar. Comparisons to a laboratory spectrophotometer demonstrate comparable results, demonstrating prototype validity.
Our initial prototype showed that using a smartphone for spectrophotometry in this particular case is feasible. However, it also revealed several limitations. Smartphone-based spectrophotometry can be prone to errors and inaccuracies based on the device's design and the experimental setup. It is important to consider the device's ease of use and the method for extracting absorbance spectra. Additionally, not every nutrient is suitable for spectrophotometric analysis.

In the second-generation prototype, we narrowed our target analyte to a single nutrient (vitamin B12) and looked to improve result quality and overall ease of use while maintaining accessibility. To do so, we implemented a more mature, portable design and a custom, semi-automatic signal analysis pipeline. This second-generation prototype was evaluated against the first-generation prototype to analyze the improvements. 
Importantly, we conducted experiments to inform spectrophotometer design parameters for more reliable results. In these tests, we found that the light source applied greatly impacted results and that an LED light bulb was the most optimal because of its even emission spectra across the visible spectrum. Conversely, the method of reference spectra calculation did not significantly impact results. Under these conditions, our smartphone-based spectrophotometer can be comparable to a laboratory spectrophotometer for identifying and quantifying vitamin B12. 


While this work takes several meaningful steps to inform future solutions, there are still limitations that must be addressed. The most pressing of these is the high variance exhibited by both prototypes relative to the laboratory device. This could be caused by outside light interfering with the analysis, a fixed and non-adjustable light slit design, and/or flaws and artifacts in the diffraction grating or the smartphone camera. Associated with these issues, we also found that the smartphone-based prototypes had an overall poor coverage of the full visible spectrum of light. Figure \ref{fig:supplement-comp} is a good example; we observe that the absorbance spectra is made up mostly of noise at wavelengths less than 450 nm or more than 650 nm. 
It seems that the reduction in the feasible detection range is likely caused by the diffraction grating. This issue could potentially be resolved by making the diffraction grating larger, positioning it more directly in the path of light, and/or substituting it with a commercial grating, although this may make the device less accessible. In terms of methodology, these experiments are limited by the use of a single type of nutrient, a single smartphone model, and the reliance on nutrient supplements (which may have additional compounds aside from the nutrient itself). 

Although the results are limited, they should encourage future research to explore smartphone-based spectrophotometry for nutrient assessment. This exploration should focus on improving device reliability, expanding applications to a wider range of nutrients, and considering real-world scenarios. We envision future iterations that can detect nutritional status through spectrophotometric blood analysis via the epidermis or through less invasive biosamples such as saliva or urine.

\section{Conclusion}
In this work, we explored smartphone-based spectrophotometry for nutrient identification and quantification. We built and evaluated two smartphone-based spectrophotometer prototypes and validated the second-generation prototype against a benchtop laboratory spectrophotometer. We further experimented with and analyzed the effects of four different light sources, as well as three different reference spectra strategies. The qualitative and quantitative results show that the second-generation device, which is integrated with a custom semi-automatic signal analysis pipeline, is an improvement over the first generation. By applying the Beer-Lambert Law, the results indicate that this device can  quantify vitamin B12 in a solution, with an accuracy ranging from 68.6\% to 91.3\%, when compared to the laboratory spectrophotometer. We thoroughly discussed the successes and limitations of the prototypes, emphasizing implications for future design iterations. This work indicates that smartphone-based spectrophotometry is a promising technology for nutrient identification and quantification and demonstrates the potential for a future device that can interface directly with the human body. We envision that with the adoption of such approaches, people, especially in resource-constrained regions, can accessibly assess their nutrient status.

\section*{Acknowledgment}
Special thanks to Kyan Yang for his help programming the signal processing pipeline.

\bibliographystyle{IEEEtranS}
\bibliography{IEEEabrv,main}
 
\end{document}